\documentclass[12pt]{article}
\usepackage{amsmath}
\begin{document}

\title{\hfill{\normalsize{}hep-th/0410215 v4}\\[5mm]
{\bf{}BRST approach to Lagrangian construction for
fermionic massless higher spin fields}}

\author{\sc I.L. Buchbinder${}^{a}$,
V.A. Krykhtin${}^b$,\footnote{e-mail:
\tt joseph@tspu.edu.ru, krykhtin@mph.phtd.tpu.edu.ru}
\fbox{A. Pashnev${}^c$}\\[0.5cm]
\it ${}^a$Department of Theoretical Physics,\\
\it Tomsk State Pedagogical University,\\
\it Tomsk 634041, Russia\\[0.3cm]
\it ${}^b$Laboratory of Mathematical Physics and\\
\it Department of Theoretical and Experimental Physics\\
\it Tomsk Polytechnic University,\\
\it Tomsk 634050, Russia\\[0.3cm]
\it ${}^c$Bogolubov Laboratory of Theoretical Physics,\\
\it  JINR, Dubna, 141980 Russia\\}

\date{}

\maketitle
\thispagestyle{empty}

\begin{abstract}
We develop the BRST approach to Lagrangian formulation
for all massless half-integer higher spin fields on an
arbitrary dimensional flat space.
General procedure of Lagrangian construction describing
the dynamics of fermionic field with any spin is given.
It is shown that in fermionic case the higher spin
field model is a reducible gauge theory and the
order of reducibility grows with the value of spin.
No off-shell constraints on the fields and the gauge
parameters are used.
We prove that in four dimensions after partial gauge
fixing the Lagrangian obtained can be transformed
to Fang-Fronsdal form however, in general case,
it includes the auxiliary fields and possesses
the more gauge symmetries in compare
with Fang-Fronsdal Lagrangian.
As an example of general procedure, we derive
the new Lagrangian for spin 5/2 field containing
all set of auxiliary fields and gauge symmetries
of free fermionic higher spin field theory.
\end{abstract}

\section{Introduction}
Construction of the self-consistent Lagrangian theory of interacting
higher spin fields is one of the longstanding problems of the
theoretical physics. First success in the theory of massless higher
spin fields was the formulation of Lagrangians for free bosonic
\cite{Fronsdal} and fermionic \cite{Fang} fields in four dimensions.
Since then the various approaches to higher spin fields problem were
developed (see e.g.  \cite{reviews} for reviews and
\cite{Francia,dev} for
recent developments) however the general problem is still open.
We would like to point out two modern approaches.

An approach, called the unfolded formalism, was
developed by Vasiliev et al.
(see e.g.  \cite{Vasiliev} and references therein). It allows to
construct both the theory of free
higher spin fields and the theory of higher spin fields coupled to
$AdS_D$ background (see \cite{Lopatin} for the bosonic case and
\cite{V-f} for the fermionic case and references therein).  Also this
formalism turned out to be fruitful for constructing the consistent
equations of motion for interacting higher spin fields.

Another approach to higher spin field problem,
called
BRST\footnote{BRST construction was discovered at first in
context of Yang-Mills theories \cite{BRST}}
approach
\cite{Bengtsson}, was initiated by development of string field theory
where the interacting model of open strings was constructed
(see e.g. \cite{Witten}) on the base
of BRST techniques\footnote{Also we point out the approach
\cite{Siegel} to finding the gauge invariant actions for
arbitrary representations of the Poincare group.}.
Higher spin BRST approach
is analogous to string field theory however it contains two essential
differences related with structure of constraints, which are used
in construction of BRST charge, and with presence only massless fields
in the spectrum of higher spin field model.
If ones try to consider the
tensionless limit of the string field theory
(see e.g. for free string theory \cite{0305155,0311257})
we expect to get the theory of
interacting massless fields.  Since the string field theory contains
lesser number of constraints on the fields than we need to construct an
irreducible representation of Poincare group, fields in string spectrum
do not belong to irreducible representations with fixed spin and
their equations of motion describe rather propagation of Regge
trajectories, instead of one spin mode.
For the equations of motion to
describe propagation of one spin mode the additional, in compare with
string, off-shell constraints on the fields must be imposed. In order
to get Lagrangian which contains the additional constraints as
equations of motion we have to include these additional constraints
into the set of constraints which is used in constructing the BRST
charge and then try to get the Lagrangian of the higher spin field
theory. Using this approach one can hope to construct the theory of
interacting higher spin fields analogously to the string field theory.
(An attempt to do that was undertaken in \cite{attempt}.)

The first natural step of constructing the massless higher spin
interacting model in the BRST approach is a formulation of
corresponding free model. This problem was studied in \cite{Bengtsson}
and finally solved for the bosonic massless higher spin fields both on
the flat \cite{9803207,0101201} and the AdS \cite{0109067,0206026}
backgrounds. However, the BRST approach to fermionic fields has not
been developed at all so far.

The present paper is devoted to formulation of BRST approach to
derivation the Lagrangian for free fermionic massless
higher spin fields on the flat Minkowski space of arbitrary dimension.
The method which we use here slightly differs from the one in the
bosonic case, but application of our method in the bosonic case leads
to the same final result for the Lagrangian describing propagation of
all spin fields.  The difference of the methods is rather technical and
consists in that we do not use the similarity transformation like in
\cite{0101201} and therefore one can construct Lagrangian for the field
of one fixed value of spin while the approach in \cite{0101201} demands
to use fields of all spins together.
As a future purpose we hope, using our method, to construct the free
theory of massive higher fermionic fields (see
e.g.  \cite{0312252} and references therein), to get
the Lagrangian describing propagation of fermionic higher spin fields
through AdS background, and to consider an application of BRST approach
to supersymmetric higher superspin models (for recent
development in these directions see e.g.  \cite{mAdS} for massive
higher spin field in AdS background and \cite{SuSy} for supersymmetric
higher spin field models).

The paper is organized as follows.
In Section~\ref{ac} we investigate the superalgebra
generated by the constraints which are necessary to define a
irreducible half-integer spin representation of Poincare group.
It is shown that an naive use of the BRST charge, constructed on
the base of these constraints, leads
us to the equations of motion only for spin-1/2 fields.
We argue, to overcome this difficulty we should reformulate the
constraint algebra and find a new representation for the constraints.

Section~\ref{auxalg} is devoted to actual formulation of new
representation for constraints.

In Section~\ref{Lagr} we construct Lagrangian
describing a propagation of field with any fixed
half-integer spin.
We find that in the case of arbitrary spin fermionic field,
the theory
has reducible gauge symmetry with the finite order of reducibility
which increases with the spin value.
Next, in Section~\ref{LagrAll} we construct Lagrangian
describing propagation of all half-integer spin fields
simultaneously.

Then in Section~\ref{s:FF} we show that in four dimensions found
Lagrangian may be transformed, after partial gauge fixing, to
the Fang~\&~Fronsdal Lagrangian \cite{Fang}.

In Section~\ref{example} we illustrate the general procedure of
Lagrangian construction by finding the
Lagrangian and gauge transformations for the spin $5/2$ field model
without gauge fixing, keeping all auxiliary fields and higher
spin gauge symmetries.

This paper is devoted to the memory of our friend and
collaborator, a remarkable human being and scientist A.I. Pashnev who
tragically passed away on March 30, 2004.

\section{Algebra of the constraints}\label{ac}

It is well known that the totally symmetrical tensor-spinor field
$\Psi_{\mu_1\cdots\mu_n}$ (the Dirac index is suppressed),
describing the irreducible spin $s=n+1/2$ representation must
satisfy the following constraints (see e.g. \cite{BK})
\begin{eqnarray}
&&
\gamma^\nu\partial_\nu \Phi_{\mu_1\cdots\mu_n}=0,
\label{irrep0}
\\
&&
\gamma^\mu\Phi_{\mu\mu_2\cdots\mu_n}=0.
\label{irrep1}
\end{eqnarray}
Here $\gamma^\mu$ are the Dirac matrices
$\{\gamma_\mu,\gamma_\nu\}=2\eta_{\mu\nu}$,
$\eta_{\mu\nu}=(+,-,\ldots,-)$.

In order to describe all higher tensor-spinor fields
together it is convenient to introduce Fock space
generated by creation and annihilation operators $a_\mu^+$,
$a_\mu$ with vector Lorentz index $\mu=0,1,2,\ldots,D-1$
satisfying the commutation relations
\begin{eqnarray}
\bigl[a_\mu,a_\nu^+\bigr]=-\eta_{\mu\nu}.
\end{eqnarray}
These operators act on states in the
Fock space
\begin{eqnarray}
|\Phi\rangle&=&\sum_{n=0}^{\infty}\Phi_{\mu_1\cdots\mu_n}(x)
a^{+\mu_1}\cdots a^{+\mu_n}|0\rangle
\label{gstate}
\end{eqnarray}
which describe all half-integer spins simultaneously if the following
constraints are taken into account
\begin{eqnarray}
T_0|\Phi\rangle&=&0,
\qquad
T_1|\Phi\rangle=0,
\label{01}
\end{eqnarray}
where $T_0=\gamma^\mu{}p_\mu$, $T_1=\gamma^\mu{}a_\mu$. If
constraints (\ref{01}) are fulfilled for the general state
(\ref{gstate}) then constraints (\ref{irrep0}), (\ref{irrep1})
are fulfilled for each component $\Phi_{\mu_1\cdots\mu_n}(x)$ in
(\ref{gstate}) and hence the relations (\ref{01}) describe all
free higher spin fermionic fields together.
In order to construct
hermitian BRST charge we have to take into account the
constraints which are hermitian conjugate to $T_0$ and $T_1$.
Since $T_0^+=T_0$ we have to add only one constraint
$T_1^+=\gamma^\mu{}a_\mu^+$ to the initial constraints $T_0$ and
$T_1$.

Algebra of the constraints $T_0$, $T_1$, $T_1^+$ is not closed
and in order to construct the BRST charge we must include in
the algebra of constraints all the constraints generated by
$T_0$, $T_1$, $T_1^+$. The resulting constraints and their
algebra are written in Table~\ref{table}.
\begin{table}[t]
\small
\begin{eqnarray*}
\begin{array}{||l||r|r|r||r|r|r|r|r||r||}\hline\hline\vphantom{\Biggm|}
&T_0&T_1&T_1^+&\quad{}L_0&L_1&L_1^+&L_2&L_2^+&G_0\\
\hline\hline\vphantom{\Biggm|}
T_0=\gamma^\mu{}p_\mu&-2L_0&2L_1&2L_1^+&0&0&0&0&0&0\\
\hline\vphantom{\Biggm|}
T_1=\gamma^\mu{}a_\mu&2L_1&4L_2&-2G_0&0&0&-T_0&0&-T_1^+&T_1\\
\hline\vphantom{\Biggm|}
T_1^+=\gamma^\mu{}a_\mu^+&2L_1^+&-2G_0&4L_2^+&0&T_0&0&T_1&0&-T_1^+\\
\hline\hline\vphantom{\Biggm|}
L_0=-p^2&0&0&0&0&0&0&0&0&0\\
\hline\vphantom{\Biggm|}
L_1=p^\mu{}a_\mu&0&0&-T_0&0&0&L_0&0&-L_1^+&L_1 \\
\hline\vphantom{\Biggm|}
L_1^+=p^\mu{}a_\mu^+&0&T_0&0&0&-L_0&0&L_1&0&-L_1^+ \\
\hline\vphantom{\Biggm|}
L_2=\frac{1}{2}a_\mu{}a^\mu&0&0&-T_1&0&0&-L_1&0&G_0&2L_2\\
\hline\vphantom{\Biggm|}
L_2^+=\frac{1}{2}a_\mu^+{}a^{\mu+}&0&T_1^+&0&0&L_1^+&0&-G_0&0&-2L_2^+\\
\hline\hline\vphantom{\Biggm|}
G_0=-a_\mu^+a^\mu+\frac{D}{2}&0&-T_1&T_1^+&0&-L_1&L_1^+&-2L_2&2L_2^+&0\\
\hline\hline
\end{array}
\end{eqnarray*}
\caption{Algebra of the constraints}\label{table}
\end{table}
The constraints $T_0$, $T_1$, $T_1^+$ are fermionic and the
constraints $L_0$, $L_1$, $L_1^+$, $L_2$, $L_2^+$, $G_0$ are
bosonic.
All the commutators are graded, i.e. graded commutators between
the fermionic constraints are anticommutators and graded
commutators which include any bosonic constraint are
commutators.
In Table~\ref{table} the first arguments of the graded
commutators and explicit expressions for all the constraints are
listed in the left column and the second argument of graded
commutators are listed in the upper row.
It is worth pointing out that this algebra involves all the bosonic
constraints $L_0$, $L_1$, $L_1^+$, $L_2$, $L_2^+$, $G_0$ which
were used to describe the irreducible bosonic representation in
\cite{9803207} as a subalgebra.

Let us introduce the BRST charge for the enlarged
system of constraints
\begin{eqnarray}
\nonumber
Q'
&=&
q_0T_0+q_1^+T_1+q_1T_1^+
+\eta_0L_0+\eta_1^+L_1+\eta_1L_1^+
+\eta_2^+L_2
+\eta_2L_2^+
+\eta_{G}G_0
\\&&
\nonumber{}
+i(\eta_1^+q_1-\eta_1q_1^+)p_0
-i(\eta_Gq_1+\eta_2q_1^+)p_1^+
+i(\eta_Gq_1^++\eta_2^+q_1)p_1
\\&&
\nonumber{}
+(q_0^2-\eta_1^+\eta_1){\cal{}P}_0
+(2q_1q_1^+-\eta_2^+\eta_2){\cal{}P}_G
\\&&
\nonumber{}
+(\eta_G\eta_1^++\eta_2^+\eta_1-2q_0q_1^+){\cal{}P}_1
+(\eta_1\eta_G+\eta_1^+\eta_2-2q_0q_1){\cal{}P}_1^+
\\&&
{}
+2(\eta_G\eta_2^+-q_1^{+2}){\cal{}P}_2
+2(\eta_2\eta_G-q_1^2){\cal{}P}_2^+.
\label{bareBRST}
\end{eqnarray}
Here $q_0$, $q_1^+$, $q_1$ are the bosonic ghosts corresponding to
the fermionic constraints $T_0$, $T_1$, $T_1^+$ respectively and
$\eta_0$, $\eta_1^+$, $\eta_1$, $\eta_2^+$, $\eta_2$, $\eta_G$
are fermionic ghosts corresponding to the bosonic constraints.
The momenta for these ghosts are $p_0$, $p_1$,
$p_1^+$ for bosonic and ${\cal{}P}_0$, ${\cal{}P}_1$,
${\cal{}P}_1^+$, ${\cal{}P}_2$, ${\cal{}P}_2^+$, ${\cal{}P}_G$
for fermionic ones. They satisfy the usual commutation relations
\begin{eqnarray}
&&
\{\eta_0,{\cal{}P}_0\}=
\{\eta_G,{\cal{}P}_G\}=
\{\eta_1,{\cal{}P}_1^+\}=
\{\eta_1^+,{\cal{}P}_1\}=
\{\eta_2,{\cal{}P}_2^+\}=
\{\eta_2^+,{\cal{}P}_2\}=1,
\\&&
[q_0,p_0]=
[q_1,p_1^+]=
[q_1^+,p_1]=i
\end{eqnarray}
and act on the vacuum state as follows
\begin{equation}
 p_0|0\rangle
=q_1|0\rangle
=p_1|0\rangle
={\cal{}P}_0|0\rangle
={\cal{}P}_G|0\rangle
=\eta_1|0\rangle
={\cal{}P}_1|0\rangle
=\eta_2|0\rangle
={\cal{}P}_2|0\rangle
=0.
\end{equation}

The BRST charge (\ref{bareBRST}) acts in enlarged space of state
vectors
depending both on $a^{+\mu}$ and on the
ghost operators $q_0$, $q_1^+$, $p_1^+$, $\eta_0$, $\eta_G$,
$\eta_1^+$, ${\cal{}P}_1^+$, $\eta_2^+$, ${\cal{}P}_2^+$ and
having the structure
\begin{eqnarray}
|\Psi\rangle
&=&
\sum_{k_i}(q_0)^{k_1} (q_1^+)^{k_2} (p_1^+)^{k_3} (\eta_0)^{k_4}
(\eta_G)^{k_5}(\eta_1^+)^{k_6} ({\cal{}P}_1^+)^{k_7}
(\eta_2^+)^{k_8} ({\cal{}P}_2^+)^{k_9}
\times
\nonumber
\\
&&\qquad{}
\times
a^{+\mu_1}\cdots a^{+\mu_{k_0}}
\Psi_{\mu_1\cdots\mu_{k_0}}^{k_1\cdots{}k_9}(x)|0\rangle
\label{state}
\end{eqnarray}
The corresponding ghost number is $0$.
The sum in (\ref{state}) is assumed over $k_0$, $k_1$, $k_2$,
$k_3$ running from 0 to infinity and over $k_4$, $k_5$, $k_6$,
$k_7$, $k_8$, $k_9$ running from 0 to 1.
It is evident that the state vectors
(\ref{gstate}) are the partial cases of the above vectors.

The physical states are defined in
the BRST approach by the equation
\begin{eqnarray}
Q'|\Psi\rangle&=&0
\end{eqnarray}
which is treated as an equation of motion. Besides,
if $|\Psi\rangle$ is a physical state, then due to nilpotency
of the BRST operator, the state $|\Psi\rangle+Q'|\Lambda\rangle$ will
also be physical for any $|\Lambda\rangle$.
It means we have the gauge
transformations
\begin{eqnarray}
\delta|\Psi\rangle&=&Q'|\Lambda\rangle.
\end{eqnarray}

Let us decompose the BRST charge $Q'$ (\ref{bareBRST}), the
state vector $|\Psi\rangle$ and the parameter of the gauge
transformations $|\Lambda\rangle$ as follows
\begin{eqnarray}
Q'&=&Q_0+\eta_G\tilde{G}_0+(2q_1^+q_1-\eta_2^+\eta_2){\cal{}P}_G,
\qquad
[\tilde{G}_0,Q_0]=0,
\\
&&
\tilde{G}=G_0-iq_1p_1^++iq_1^+p_1
+\eta_1^+{\cal{}P}_1-\eta_1{\cal{}P}_1^+
+2\eta_2^+{\cal{}P}_2-2\eta_2{\cal{}P}_2^+,
\\
|\Psi\rangle&=&|\Psi_0\rangle+\eta_G|\Psi_G\rangle,
\\
|\Lambda\rangle&=&|\Lambda_0\rangle+\eta_G|\Lambda_G\rangle.
\end{eqnarray}
Here the state vectors $|\Psi_0\rangle$, $|\Psi_G\rangle$ and
the gauge parameters $|\Lambda_0\rangle$, $|\Lambda_G\rangle$
are independent of $\eta_G$.
Then the equations of motion and the gauge transformations take the
form
\begin{eqnarray}
&&
Q_0|\Psi_0\rangle+(2q_1^+q_1-\eta_2^+\eta_2)|\Psi_G\rangle=0,
\label{0Psi}
\\
&&
\tilde{G}_0|\Psi_0\rangle-Q_0|\Psi_G\rangle=0,
\label{GPsi}
\\[2mm]
&&
\delta|\Psi_0\rangle=Q_0|\Lambda_0\rangle
 +(q_1^+q_1-\eta_2^+\eta_2)|\Lambda_G\rangle,
\\
&&
\delta|\Psi_G\rangle=\tilde{G}_0|\Lambda_0\rangle-Q_0|\Lambda_G\rangle.
\end{eqnarray}

Now ones try to simplify these equations.
First, we decompose the state vector $|\Psi\rangle$ and the gauge
parameter $|\Lambda\rangle$ in the eigenvectors of operator
$\tilde{G}_0$:
$|\Psi\rangle=\sum|\Psi_n\rangle$,
$|\Lambda\rangle=\sum|\Lambda_n\rangle$,
with
$\tilde{G}_0|\Psi_n\rangle=(n+\frac{D-4}{2})|\Psi_n\rangle$,
$\tilde{G}_0|\Lambda_n\rangle=(n+\frac{D-4}{2})|\Lambda_n\rangle$,
$n=0,1,2,\ldots$.
Then using the gauge transformation we can make all
$|\Psi_{Gn}\rangle=0$ choosing
$|\Lambda_n\rangle=\frac{-1}{n+(D-4)/2}|\Psi_G\rangle$
except the case $n+\frac{D-4}{2}=0$.
When $D=4$ we
have $n=0$ and the field $|\Psi_G\rangle$ after this gauge
transformation is reduced to
\begin{eqnarray}
|\Psi_G\rangle&\to&|\Psi_{G0}\rangle=\psi_G(x)|0\rangle,
\label{psiG}
\end{eqnarray}
i.e. it contains only spin-$1/2$ field.
Substituting (\ref{psiG}) in the equations of motion
(\ref{0Psi}), (\ref{GPsi}) ones get
\begin{eqnarray}
&&
\sum_{n=0}^{\infty}Q_0|\Psi_{0n}\rangle=0,
\\
&&
\sum_{n=0}^{\infty}
\Bigl(n+\frac{D-4}{2}\Bigr)|\Psi_{0n}\rangle
=Q_0|\Psi_{G0}\rangle.
\label{*}
\end{eqnarray}
Acting by the operator $\tilde{G}_0$ on both sides of equation
(\ref{*}) ones obtain
\begin{eqnarray}
\sum_{n=0}^{\infty}
\Bigl(n+\frac{D-4}{2}\Bigr)^2|\Psi_{0n}\rangle
=0.
\label{**}
\end{eqnarray}
Since all the states vectors
$|\Psi_{0n}\rangle$ are linear independent, equation (\ref{**})
means that all $|\Psi_{0n}\rangle=0$, except
$n=-\frac{D-4}{2}=0$. Thus analogously to (\ref{psiG}) one can
write
\begin{eqnarray}
|\Psi_0\rangle&\to&|\Psi_{00}\rangle=\psi(x)|0\rangle.
\end{eqnarray}
Ultimately we have two independent equations of motion
\begin{eqnarray}
&&
T_0|\Psi_{00}\rangle=0,
\qquad
T_0|\Psi_{G0}\rangle=0
\label{doubling}
\end{eqnarray}
both for $|\Psi_{00}\rangle$ and $|\Psi_{G0}\rangle$.
These equations in component form read
\begin{eqnarray}
\gamma^\mu p_\mu\psi(x)=0,
\qquad
\gamma^\mu p_\mu\psi_G(x)=0.
\end{eqnarray}
So, we see that the above construction leads to
double number of equations and only for
spin-1/2 fields.
Hence, such a procedure is unsatisfactory.

To clarify a situation we pay attention to two points.

First,
if we suppose that the state vectors and the gauge parameters do
not depend on the ghost field $\eta_G$ then we have only one
Dirac equation and avoid the doubling the physical component states.

Second,
the above construction has led us to the equations only for
spin-1/2 fields.
This happens because of $\tilde{G}_0$ has the structure
\begin{eqnarray}
\tilde{G}_0&=&\tilde{N}_0+\frac{D-4}{2}\,,
\end{eqnarray}
where $\tilde{N}_0$ is proportional to the "particle" number operators
associated with the operators
$a^{+\mu}$, $q_1^+$, $p_1^+$, $\eta_1^+$, ${\cal{}P}_1^+$,
$\eta_2^+$, ${\cal{}P}_2^+$
and
therefore if we want $G_0$ to be  considered as a
constraint,
we get that
there
are no "particles" (in $D=4$ case)
in the physical states
and hence only equations of
motion for the field with spin 1/2 arise.

We note that if we had instead of
constraint $\tilde{G}$ another constraint $\tilde{G}_0+h$ with $h$
being an arbitrary constant, we could get equations of motion
for fields of any spin by choosing the arbitrary parameter $h$ in the
proper way for each spin. As a result, we could put $n$ to any integer
number since instead of condition $n+\frac{D-4}{2}=0$ we had
condition $n+\frac{D-4}{2}+h=0$.  However, if we simply change the
constraint $G_0\to{}G_0+h$ we break the algebra of the constraints
which is given in Table~\ref{table}.  Thus, introducing of this
arbitrary parameter $h$ must be carry out in such a way that the
algebra of the constraints will not be broken. This discussion shows
that the representation for the constraints we used is too naive
and improper and we have to find another representation.

Such a new representation may be realized as follows.
We enlarge the number of creation and annihilation operators and
extend the
expressions for the constraints using the prescription:
the new expressions for the constraints should have the general
structure
\begin{eqnarray}
\mbox{\it{}New constraint}&=& \mbox{\it{}Old
constraint}+\mbox{\it{}Additional part},
\label{newstruct}
\end{eqnarray}
with some additional parts which will be found in Section 3 in explicit
form.
This new representation
must be constructed in such a way that the arbitrary parameter
$h$ appears in constraint $G_{0new}$ as follows
\begin{eqnarray}
G_{0new}&=&G_0+N_{add}+C+h, \label{Gstructure}
\end{eqnarray}
where $N_{add}$ is proportional to the number operators
of additional "particles", associated with extra annihilation and
creation operators, $C$ is some fixed constant which may arise.

It is evident that we must construct these additional parts
first of all for those
constraints whose commutators give $G_0$.
The corresponding operators are
$T_1$, $T_1^+$, $L_2$, $L_2^+$.
Then we must go on and construct additional parts for those
constraints whose commutators give
$T_1$, $T_1^+$, $L_2$, $L_2^+$.
Fortunately, they are
$G_0$, $T_1$, $T_1^+$, $L_2$, $L_2^+$
and we may construct no additional parts for the other
operators.
Note that these operators
form a subalgebra.
Thus we
have to construct a representation for the subalgebra of the
constraints $G_0$, $T_1$, $T_1^+$, $L_2$, $L_2^+$ only.
In the next Section we describe construction
of a representation
for such an subalgebra.

Certainly, we can construct additional parts for all
the operators of the algebra given in Table~\ref{table}, but in
this case we must use some massive parameter
which is absent in the true massless theory.
Of course, one can try to introduce such a massive parameter
to constraints by hand. However in this case we expect to get
a massive higher spin field theory
\cite{xxx}.

\section{New representation of the constraints}\label{auxalg}

In this Section we construct a new representation for the
algebra of the constraints
so that the new expressions for the constraints have the structure
(\ref{newstruct}) and the parameter $h$ appears in the
constraint $G_0$ in the proper way (\ref{Gstructure}).
Algebra of the new constraints still has the form given by
Table~\ref{table}.
As was explained at the end of the previous Section,
for this purpose
it is enough
to construct the additional parts only for
$G_0$, $T_1$, $T_1^+$, $L_2$, $L_2^+$
and the new expressions for the constraints should be
\begin{align}
\label{add1}
L_{2new}^+&=L_2^++L_{2add}^+,&L_{2new}&=L_2+L_{2add},\\
T_{1new}^+&=T_1^++T_{1add}^+,&T_{1new}&=T_1+T_{1add},\\
G_{0new}&=G_0+G_{0add}
\label{add3}
\end{align}
(all the other constraints do not change).
Since
the constraints
$G_{0new}$, $T_{1new}$, $T_{1new}^+$, $L_{2new}$, $L_{2new}^+$
form a subalgebra and since the old and additional expressions
of the constraints commute,
the
$G_{0add}$, $T_{1add}$, $T_{1add}^+$, $L_{2add}$, $L_{2add}^+$
form a subalgebra too with the same commutation relation
among them as for the old expressions for the constraints
$G_0$, $T_1$, $T_1^+$, $L_2$, $L_2^+$.
Thus
it is enough to find a representation of the subalgebra in terms
of new creation and annihilation operators which will be
introduced later.

Let us turn to the construction of the subalgebra
representation.

Note that the commutation relations between $G_0$ and the other
operators of the subalgebra resemble the commutation relations
between a number operator and creation and
annihilation operators.
Therefore
let us consider the representation of the subalgebra of the
constraints
with the state vector $|0\rangle_V$
annihilated by the operators $T_1$ and $L_2$
\begin{eqnarray}
T_1|0\rangle_V=L_2|0\rangle_V=0
\label{V1}
\end{eqnarray}
and being the eigenvector of the operator $G_0$
\begin{eqnarray}
G_0|0\rangle_V=h|0\rangle_V,
\label{V2}
\end{eqnarray}
where $h$ is an arbitrary constant.\footnote{The representation
which is given by (\ref{V1}) and (\ref{V2})  is called in the
mathematical literature the Verma module. It explains the
subscript $V$ at the state vectors.}
It is
the relation (\ref{V2}) where "the vacuum state" $|0\rangle_V$
is an eigenvector of "the number operator" $G_0$ with eigenvalue
$h$ gives us the desired structure of the operator $G_{0new}$
(\ref{Gstructure}).
Since
$(T_1^+)^2=2L_2^+$ we can choose the basis vectors in this
representation as follows
\begin{eqnarray}
|0,m\rangle_V&=&(L_2^+)^m|0\rangle_V,
\qquad
|1,m\rangle_V=T_1^+(L_2^+)^m|0\rangle_V.
\label{bv}
\end{eqnarray}

The next step is to find the action of the operators $T_1$,
$T_1^+$, $L_2$, $L_2^+$, $G_0$ on the basis vectors (\ref{bv}).
The result is
\begin{align}
\label{first}
L_2^+|0,n\rangle_V&=|0,n+1\rangle_V,&
L_2^+|1,n\rangle_V&=|1,n+1\rangle_V,
\\
L_2|0,n\rangle_V&=(n^2-n+nh)|0,n-1\rangle_V,&
L_2|1,n\rangle_V&=(n^2+nh)|1,n-1\rangle_V,
\\
T_1^+|0,n\rangle_V&=|1,n\rangle_V,&
T_1^+|1,n\rangle_V&=2|0,n+1\rangle_V,
\\
T_1|0,n\rangle_V&=-n|1,n-1\rangle_V,&
T_1|1,n\rangle_V&=-2(n+h)|0,n\rangle_V,
\\
G_0|0,n\rangle_V&=(2n+h)|0,n\rangle_V,&
G_0|1,n\rangle_V&=(2n+1+h)|1,n\rangle_V.
\label{last}
\end{align}

Now, in order to construct the new representation for the
subalgebra ones introduce the
additional creation and annihilation operators.
The number of
pairs of these operators is equal to the number of the
mutually conjugate pairs of the constraints.
So we introduce one pair of fermionic
$d^+$, $d$
(corresponding to the constraints $T_1^+$, $T_1$)
and one pair of bosonic $b^+$, $b$
(corresponding to the constraints $L_2^+$, $L_2$)
creation and
annihilation operators with the standard commutation relations
\begin{eqnarray}
\bigl\{\,d,d^+\bigr\}=1
,
\qquad\qquad
\bigl[\,b,b^+\bigr]=1
.
\end{eqnarray}
Making use of the map of the basis vectors (\ref{bv}) and the
basis vectors of the Fock space of the operators $d^+$, $b^+$
\begin{eqnarray}
|0,n\rangle_V\longleftrightarrow(b^+)^n|0\rangle\equiv|n\rangle,
\qquad
|1,n\rangle_V\longleftrightarrow d^+|n\rangle
\label{map}
\end{eqnarray}
we can construct a representation of the subalgebra.
>From (\ref{first})--(\ref{last}) and (\ref{map}) ones find
\begin{align}
\label{firsto}
L_{2add}^+&=b^+,&L_{2add}&=(b^+b+d^+d+h)b,\\
\label{lasto}
T_{1add}^+&=2b^+d+d^+,&T_{1add}&=-2(b^+b+h)d-d^+b,\\
G_{0add}&=2b^+b+d^+d+h.
\label{Gadd}
\end{align}
It is easy to see, the operators
(\ref{firsto}), (\ref{lasto}) are not hermitian conjugate to each
other
\begin{eqnarray}
(T_{1add})^+\neq T_{1add}^+,
&\qquad&
(L_{2add})^+\neq L_{2add}^+
\end{eqnarray}
if we use the usual rules for hermitian conjugation of the
additional creation and annihilation operators
\begin{eqnarray}
(d)^+=d^+,&\qquad&(b)^+=b^+.
\end{eqnarray}

The reason is that the map (\ref{map}) does not preserve the scalar
product. If we have two vectors
$|\Phi_1\rangle_V$ and $|\Phi_2\rangle_V$ and corresponding them
two vectors in the Fock space
$|\Phi_1\rangle\leftrightarrow|\Phi_1\rangle_V$,
$|\Phi_2\rangle\leftrightarrow|\Phi_2\rangle_V$ then in general
\begin{eqnarray}
{}_V\langle\Phi_1|\Phi_2\rangle_V
\neq
\langle\Phi_1|\Phi_2\rangle,
\end{eqnarray}
where we assumed that ${}_V\langle0|0\rangle_V=1$.
In order to improve the situation we change the scalar product
in the Fock space so that
\begin{eqnarray}
{}_V\langle\Phi_1|\Phi_2\rangle_V
=
\langle\Phi_1|\Phi_2\rangle_{new}
=
\langle\Phi_1|K|\Phi_2\rangle,
\label{proof}
\end{eqnarray}
with some operator $K$.

This operator may be found as follows. If we have a map between
two bases
\begin{eqnarray}
|n\rangle_V\longleftrightarrow|n\rangle,
\end{eqnarray}
then
\begin{eqnarray}
|\Phi\rangle_V=\sum c_n|n\rangle_V
\longleftrightarrow
\sum c_n|n\rangle=|\Phi\rangle.
\label{scalar}
\end{eqnarray}
Therefore if we preserve the scalar product for the basis
vectors ${}_V\langle{}m|n\rangle_V=\langle{}m|K|n\rangle$ then
it will be preserved for all vectors.

In the case of orthogonal basis in the Fock space,
$\langle{}m|n\rangle=C_n\delta_{mn}$ with $C_n$ being some
constants (as we have in our case)
it may be proved by direct substitution to (\ref{proof}) that
the operator $K$ is
\begin{eqnarray}
K&=&\sum
|m\rangle\frac{{}_V\langle{}m|n\rangle_V}{C_mC_n}\,\langle{}n|.
\end{eqnarray}
Hence, in the case under consideration we get
\begin{eqnarray}
\label{K}
K&=&
\sum_{n=0}^\infty\frac{1}{n!}
  \Bigl(\,
     |n\rangle{}\langle{}n|\,C(n,h)
     -
     2d^+|n\rangle\langle{}n|d\,C(n+1,h)\,
  \,\Bigr),
\\&&
C(n,h)=\prod_{k=0}^{n-1}(k+h),
\qquad
C(0,h)=1.
\end{eqnarray}
It is a simple exercise to check that operators (\ref{firsto})
and (\ref{lasto}) are now mutually conjugate in the following sense
\begin{align}
&KT_{1add}=(T_{1add}^+)^+K,
&KT_{1add}^+=(T_{1add})^+K,
\\
&KL_{2add}=(L_{2add}^+)^+K,
&KL_{2add}^+=(L_{2add})^+K.
\end{align}

Thus we have found the new representation of the algebra of
constraints which is given by
(\ref{add1})--(\ref{add3})
with
(\ref{firsto})--(\ref{Gadd}).
(Remind that all the other constraints of the algebra do not
change.)
Since the algebra of the constraints have not been changed, a
new BRST charge is constructed substituting
in (\ref{bareBRST}) the new constraints
instead of old.
As a result ones get
\begin{eqnarray}
\nonumber
\tilde{Q}
&=&
q_0T_0
+
q_1^+\bigl(T_1-2(b^+b+h)d-bd^+\bigr)
+
q_1\bigl(T_1^++2b^+d+d^+\bigr)
\\&&{}
\nonumber{}
+\eta_0L_0+\eta_1^+L_1+\eta_1L_1^+
+\eta_2^+\bigl(L_2+(b^+b+h+d^+d)b\bigr)
+\eta_2\bigl(L_2^++b^+\bigr)
\\&&
\nonumber
\qquad{}
+\eta_{G}\bigl(G_0+2b^+b+d^+d+h\bigr)
\\&&
\nonumber{}
+i(\eta_1^+q_1-\eta_1q_1^+)p_0
-i(\eta_Gq_1+\eta_2q_1^+)p_1^+
+i(\eta_Gq_1^++\eta_2^+q_1)p_1
\\&&
\nonumber{}
+(q_0^2-\eta_1^+\eta_1){\cal{}P}_0
+(2q_1q_1^+-\eta_2^+\eta_2){\cal{}P}_G
\\&&
\nonumber{}
+(\eta_G\eta_1^++\eta_2^+\eta_1-2q_0q_1^+){\cal{}P}_1
+(\eta_1\eta_G+\eta_1^+\eta_2-2q_0q_1){\cal{}P}_1^+
\\&&{}
+2(\eta_G\eta_2^+-q_1^{+2}){\cal{}P}_2
+2(\eta_2\eta_G-q_1^2){\cal{}P}_2^+.
\label{auxBRST}
\end{eqnarray}
Let us notice that the new BRST charge (\ref{auxBRST}) is
selfconjugate in the following sense
\begin{eqnarray}
\tilde{Q}^+K&=&K\tilde{Q},
\end{eqnarray}
with operator $K$ (\ref{K}).
Now we turn to the construction of the Lagrangians for free
fermionic higher spin fields.

\section{Lagrangians for the free fermionic fields of single
spin}\label{Lagr}

In this Section we construct the Lagrangians for free higher
spin fermionic fields using the BRST charge (\ref{auxBRST}).
Unlike the bosonic case \cite{0101201,0109067} we use here
slightly another procedure.

First, let us extract the dependence of
the new BRST charge (\ref{auxBRST}) on the ghosts $\eta_G$,
${\cal{}P}_G$
\begin{eqnarray}
\tilde{Q}&=&Q+\eta_G(\pi+h)+(2q_1^+q_1-\eta_2^+\eta_2){\cal{}P}_G,
\label{auxBRST2}
\\
&&
Q^2=(\eta_2^+\eta_2-2q_1^+q_1)(\pi+h),
\qquad
[Q,\pi]=0
\label{0potent}
\end{eqnarray}
with
\begin{eqnarray}
\pi&=&G_0+2b^+b+d^+d-iq_1p_1^++iq_1^+p_1
+\eta_1^+{\cal{}P}_1-\eta_1{\cal{}P}_1^+
+2\eta_2^+{\cal{}P}_2-2\eta_2{\cal{}P}_2^+,
\label{pi}
\\
\nonumber
Q&=&
q_0(T_0-2q_1^+{\cal{}P}_1-2q_1{\cal{}P}_1^+)
+i(\eta_1^+q_1-\eta_1q_1^+)p_0
+\eta_0L_0
+(q_0^2-\eta_1^+\eta_1){\cal{}P}_0
\\&&
\nonumber{}
+
q_1^+\bigl(T_1-2(b^+b+h)d-bd^+\bigr)
+
q_1\bigl(T_1^++2b^+d+d^+\bigr)
\\&&{}
\nonumber{}
+\eta_1^+L_1+\eta_1L_1^+
+\eta_2^+\bigl(L_2+(b^+b+h+d^+d)b\bigr)
+\eta_2\bigl(L_2^++b^+\bigr)
\\&&
-i\eta_2q_1^+p_1^+
+i\eta_2^+q_1p_1
+\eta_2^+\eta_1{\cal{}P}_1
+\eta_1^+\eta_2{\cal{}P}_1^+
-2q_1^{+2}{\cal{}P}_2
-2q_1^2{\cal{}P}_2^+.
\label{Q}
\end{eqnarray}

Second,
in order to avoid the doubling the physical component states as it was
in Section~\ref{ac}, eq.~(\ref{doubling})
we suppose that the state vectors are independent of $\eta_G$,
i.e. ${\cal{}P}_G|\chi\rangle=0$.
Now
the general structure of the states
looks like
\begin{eqnarray}
|\chi\rangle
&=&
\sum_{k_i}(q_0)^{k_1} (q_1^+)^{k_2} (p_1^+)^{k_3} (\eta_0)^{k_4}
(d^+)^{k_5}(\eta_1^+)^{k_6} ({\cal{}P}_1^+)^{k_7}
(\eta_2^+)^{k_8} ({\cal{}P}_2^+)^{k_9}(b^+)^{k_{10}}
\times
\nonumber
\\
&&\qquad{}
\times
a^{+\mu_1}\cdots a^{+\mu_{k_0}}
\chi_{\mu_1\cdots\mu_{k_0}}^{k_1\cdots{}k_{10}}(x)|0\rangle.
\label{chistate}
\end{eqnarray}
The corresponding ghost number is $0$, as usual.
The sum in (\ref{chistate}) is assumed over $k_0$, $k_1$, $k_2$,
$k_3$, $k_{10}$ running from 0 to infinity and over $k_4$, $k_5$, $k_6$,
$k_7$, $k_8$, $k_9$ running from 0 to 1.

After this assumption
the equation on the physical states in the BRST approach
$\tilde{Q}|\chi\rangle=0$ yields two equations
\begin{eqnarray}
\label{Qchi}
Q|\chi\rangle&=&0,
\\
(\pi+h)|\chi\rangle&=&0.
\label{eigenv}
\end{eqnarray}
Equation~(\ref{eigenv}) is the eigenvalue equation
for the operator $\pi$ (\ref{pi}) with the corresponding
eigenvalues $-h$
\begin{eqnarray}
-h&=&n+\frac{D-4}{2},
\qquad
n=0,1,2,\ldots\ .
\label{n}
\end{eqnarray}
The numbers $n$ are related with the spin $s$ of the corresponding
eigenvectors as $s=n+1/2$.
Let us denote the eigenvectors of the operator $\pi$
corresponding to the eigenvalues $n+\frac{D-4}{2}$ as
$|\chi\rangle_n$
\begin{eqnarray}
\pi|\chi\rangle_n&=&\left(n+\frac{D-4}{2}\right)|\chi\rangle_n.
\label{chin}
\end{eqnarray}
Then solutions to the system of equations (\ref{Qchi}),
(\ref{eigenv}) are enumerated by $n=0,1,2,\ldots$ and satisfy
the equations
\begin{eqnarray}
Q_n|\chi\rangle_n&=&0,
\label{Qchin}
\end{eqnarray}
where in the BRST charge (\ref{Q}) we substitute
$n+\frac{D-4}{2}$ instead of $-h$ for each given equation on
spin $s=n+1/2$ field.
Thus we get that the BRST charge depends on $n$
\begin{eqnarray}
\nonumber
Q_n&=&
q_0(T_0-2q_1^+{\cal{}P}_1-2q_1{\cal{}P}_1^+)
+i(\eta_1^+q_1-\eta_1q_1^+)p_0
+\eta_0L_0
+(q_0^2-\eta_1^+\eta_1){\cal{}P}_0
\\&&
\nonumber{}
+
q_1^+\bigl(T_1-2b^+bd-bd^+\bigr)
+
q_1\bigl(T_1^++2b^+d+d^+\bigr)
\\&&{}
\nonumber{}
+\eta_1^+L_1+\eta_1L_1^+
+\eta_2^+\bigl(L_2+(b^+b+d^+d)b\bigr)
+\eta_2\bigl(L_2^++b^+\bigr)
\\&&{}
\nonumber
-i\eta_2q_1^+p_1^+
+i\eta_2^+q_1p_1
+\eta_2^+\eta_1{\cal{}P}_1
+\eta_1^+\eta_2{\cal{}P}_1^+
-2q_1^{+2}{\cal{}P}_2
-2q_1^2{\cal{}P}_2^+
\\&&{}
+(2q_1^+d-\eta_2^+b)(n+\frac{D-4}{2})
.
\label{Qn}
\end{eqnarray}

Let us rewrite the operators $Q_n$ (\ref{Qn}) in the form
independent of $n$.
This may be done by replacing $n+\frac{D-4}{2}$ in (\ref{Qn}) by
the operator $\pi$ (\ref{pi}). Then we obtain
\begin{eqnarray}
\nonumber
Q_{\pi}&=&
q_0(T_0-2q_1^+{\cal{}P}_1-2q_1{\cal{}P}_1^+)
+i(\eta_1^+q_1-\eta_1q_1^+)p_0
+\eta_0L_0
+(q_0^2-\eta_1^+\eta_1){\cal{}P}_0
\\&&
\nonumber{}
+
q_1^+\bigl(T_1-2b^+bd-bd^+\bigr)
+
q_1\bigl(T_1^++2b^+d+d^+\bigr)
\\&&{}
\nonumber{}
+\eta_1^+L_1+\eta_1L_1^+
+\eta_2^+\bigl(L_2+(b^+b+d^+d)b\bigr)
+\eta_2\bigl(L_2^++b^+\bigr)
\\&&{}
\nonumber
-i\eta_2q_1^+p_1^+
+i\eta_2^+q_1p_1
+\eta_2^+\eta_1{\cal{}P}_1
+\eta_1^+\eta_2{\cal{}P}_1^+
-2q_1^{+2}{\cal{}P}_2
-2q_1^2{\cal{}P}_2^+
\\&&{}
+(2q_1^+d-\eta_2^+b)\pi
,
\label{Qpi}
\end{eqnarray}
where
\begin{math}
Q_\pi=Q_n|_{n+\frac{D-4}{2}\to\pi}.
\end{math}
Operator (\ref{Qpi}) analogous to the BRST operator which
obtained in the bosonic case \cite{0101201,0109067} after
the dependence on the ghost fields $\eta_G$, ${\cal{}P}_G$ was
removed. Let us note that operator $Q_\pi$ is nilpotent
identically.

Now we can rewrite the set of equations (\ref{Qchin}) in the
equivalent form as one equation for all half-integer spins.
Since the operators $Q_{\pi}$ and $\pi$ commute then vectors
$Q_{\pi}|\chi\rangle_n$ belong to different eigenvalues of the
operator $\pi$ and consequently are linear independent.
Therefore we may write the set of equations (\ref{Qchin}) as one
equation summing them
\begin{eqnarray}
\sum_{n=0}^{\infty}Q_{\pi}|\chi\rangle_n
&=&
Q_{\pi}\sum_{n=0}^{\infty}|\chi\rangle_n
=
Q_{\pi}|\chi\rangle=0,
\end{eqnarray}
where we denote
\begin{eqnarray}
|\chi\rangle&=&
\sum_{n=0}^{\infty}|\chi\rangle_n.
\label{allchin}
\end{eqnarray}
Thus we obtain that the equation
\begin{eqnarray}
Q_{\pi}|\chi\rangle&=&0
\label{EMall}
\end{eqnarray}
with $|\chi\rangle$ defined by (\ref{allchin})
describes propagation of all half-integer spin fields
simultaneously.

Let us turn to the gauge transformations.
Analogously we suppose that the parameters of the gauge
transformations are also independent of $\eta_G$.
Due to eq.~(\ref{0potent})
we
have the following tower of the gauge transformations and the
corresponding eigenvalue equations for the gauge
parameters
\begin{align}
\label{76}
&\delta|\chi\rangle=Q|\Lambda\rangle,
&&(\pi+h)|\Lambda\rangle=0,
\\
&\delta|\Lambda\rangle=Q|\Lambda^{(1)}\rangle,
&&(\pi+h)|\Lambda^{(1)}\rangle=0,
\\
&\delta|\Lambda^{(i)}\rangle=Q|\Lambda^{(i+1)}\rangle,
&&(\pi+h)|\Lambda^{(i+1)}\rangle=0
\label{78}
\end{align}
where $h$ has already been determined for each spin.
Since the ghost number of the gauge parameters is reduced with
the stage of reducibility $gh(|\Lambda^{(i)}\rangle)=-(i+1)$ we
get that for each $n$ (and for the spin $s=n+1/2$ respectively) the
tower of the gauge transformations must be finite.
Thus in case of fermionic higher spin fields we have gauge symmetry with
reducible generators.

Doing the same procedure as above for the equations of motion we
may write the gauge transformations (\ref{76})--(\ref{78}) for
each given spin
\begin{align}
&\delta|\chi\rangle_n=Q_n|\Lambda\rangle_n,
&&\pi|\Lambda\rangle_n
   =\Bigl(n+\frac{D-4}{2}\Bigr)|\Lambda\rangle_n,
\\
&\delta|\Lambda\rangle_n=Q_n|\Lambda^{(1)}\rangle_n,
&&\pi|\Lambda^{(i)}\rangle_n
   =\Bigl(n+\frac{D-4}{2}\Bigr)|\Lambda^{(i)}\rangle_n,
\\
&\delta|\Lambda^{(i)}\rangle_n=Q_n|\Lambda^{(i+1)}\rangle_n
\end{align}
and for all half-integer spins simultaneously
\begin{align}
&\delta|\chi\rangle=Q_{\pi}|\Lambda\rangle,
&&|\Lambda\rangle=\sum_{n=0}^{\infty}|\Lambda\rangle_n,
\\
&\delta|\Lambda\rangle=Q_{\pi}|\Lambda^{(1)}\rangle,
\\
&\delta|\Lambda^{(i)}\rangle=Q_{\pi}|\Lambda^{(i+1)}\rangle,
&&|\Lambda^{(i)}\rangle=\sum_{n=0}^{\infty}|\Lambda^{(i)}\rangle_n.
\end{align}

Next step is to extract the zero ghost mode from the
operator $Q_{\pi}$ (\ref{Qpi}).
This operator has the structure
\begin{equation}
\label{0Q}
Q_{\pi}=
\eta_0L_0
+(q_0^2-\eta_1^+\eta_1){\cal{}P}_0
+q_0(T_0-2q_1^+{\cal{}P}_1-2q_1{\cal{}P}_1^+)
+i(\eta_1^+q_1-\eta_1q_1^+)p_0
+\Delta{}Q_{\pi},
\end{equation}
where $\Delta{}Q_{\pi}$ is independent of
$\eta_0$, ${\cal{}P}_0$, $q_0$, $p_0$.
Also we may decompose the state vector and the gauge parameters as
\begin{align}
\label{0chi}
|\chi\rangle
&=\sum_{k=0}^{\infty}q_0^k(
|\chi_0^k\rangle
+\eta_0|\chi_1^k\rangle),
&
&gh(|\chi^{k}_{m}\rangle)=-(m+k),
\\
\label{0L}
|\Lambda^{(i)}\rangle
&=\sum_{k=0}^{\infty}q_0^k(|\Lambda^{(i)}{}^k_0\rangle
+\eta_0|\Lambda^{(i)}{}^k_1\rangle),
&
&gh(|\Lambda^{(i)}{}^k_m\rangle)=-(i+k+m+1)
.
\end{align}
Following the procedure described in \cite{0311257}
we get rid of all the
fields except two $|\chi^0_0\rangle$, $|\chi^1_0\rangle$ and
equation (\ref{EMall}) is reduced to
\begin{eqnarray}
&&
\Delta{}Q_{\pi}|\chi^{0}_{0}\rangle
+\frac{1}{2}\bigl\{\tilde{T}_0,\eta_1^+\eta_1\bigr\}
|\chi^{1}_{0}\rangle
=0,
\label{EofM1all}
\\&&
\tilde{T}_0|\chi^{0}_{0}\rangle
+
\Delta{}Q_{\pi}|\chi^{1}_{0}\rangle
=0,
\label{EofM2all}
\end{eqnarray}
where $\tilde{T}_0=T_0-2q_1^+{\cal{}P}_1-2q_1{\cal{}P}_1^+$,
and $\{A,B\}=AB+BA$.

To be complete we show how equations (\ref{EofM1all}) and
(\ref{EofM2all}) can be derived from the (\ref{EMall}). First we extract
the zero ghost modes from  the BRST charge $Q_{\pi}$ (\ref{0Q}), the
state vector $|\chi\rangle$ (\ref{0chi}) and the gauge parameter
$|\Lambda\rangle$ (\ref{0L}).
After this the gauge transformation for the fields
$|\chi^k_0\rangle$, $k\ge2$ are
\begin{eqnarray}
\delta|\chi^{k}_{0}\rangle
&=&
\Delta{}Q_{\pi}|\Lambda^{k}_{0}\rangle
-\eta_1^+\eta_1|\Lambda^{k}_{1}\rangle
+(k+1)(\eta_1^+q_1-\eta_1q_1^+)|\Lambda^{k+1}_{0}\rangle
+\tilde{T}_0|\Lambda^{k-1}_{0}\rangle
+|\Lambda^{k-2}_{1}\rangle.
\end{eqnarray}
We see that we can make all fields $|\chi^k_0\rangle$,
$k\ge2$ to be zero using the gauge parameters
$|\Lambda^{k}_{1}\rangle$.

Second step is to consider the equations of motion at
coefficients $(q_0)^k$, $k\ge3$. Taking into account that all
fields $|\chi^k_0\rangle=0$, $k\ge2$, these equations are reduced
to
\begin{eqnarray}
|\chi^{k-2}_{1}\rangle=\eta_1^+\eta_1|\chi^{k}_{1}\rangle,
&\qquad&
k\ge3
\end{eqnarray}
and we find that all $|\chi^{k}_{1}\rangle=0$, $k\ge1$.

Finally we consider the equation at coefficient $(q_0)^2$ and
express $|\chi^{0}_{1}\rangle$ field from $|\chi^{1}_{0}\rangle$
\begin{eqnarray}
|\chi^{0}_{1}\rangle&=&-\tilde{T}_0|\chi^{1}_{0}\rangle.
\end{eqnarray}
So it remained only two fields $|\chi^{0}_{0}\rangle$ and
$|\chi^{1}_{0}\rangle$ and the independent equations of motion
for them are (\ref{EofM1all}) and (\ref{EofM2all}).

Since the operators $Q_{\pi}$, $\tilde{T}_0$,
$\bigl\{\tilde{T}_0,\eta_1^+\eta_1\bigr\}$ commute with the
operator $\pi$, then from equations (\ref{EofM1all}),
(\ref{EofM2all}) we may get equations of motion for fixed spin
fields
\begin{eqnarray}
&&
\Delta{}Q_{\pi}|\chi^{0}_{0}\rangle_n
+\frac{1}{2}\bigl\{\tilde{T}_0,\eta_1^+\eta_1\bigr\}
|\chi^{1}_{0}\rangle_n
=0,
\label{EofM1}
\\&&
\tilde{T}_0|\chi^{0}_{0}\rangle_n
+
\Delta{}Q_{\pi}|\chi^{1}_{0}\rangle_n
=0,
\label{EofM2}
\end{eqnarray}

These field equations
can be
deduced from the following Lagrangian\footnote{The Lagrangian is
defined up as usual to an overall factor}
\begin{eqnarray}
{\cal{}L}_n
&=&
{}_n\langle\chi^{0}_{0}|K_n\tilde{T}_0|\chi^{0}_{0}\rangle_n
+
\frac{1}{2}\,{}_n\langle\chi^{1}_{0}|K_n\bigl\{
   \tilde{T}_0,\eta_1^+\eta_1\bigr\}|\chi^{1}_{0}\rangle_n
\nonumber
\\&&\qquad{}
+
{}_n\langle\chi^{0}_{0}|K_n\Delta{}Q_{\pi}|\chi^{1}_{0}\rangle_n
+
{}_n\langle\chi^{1}_{0}|K_n\Delta{}Q_{\pi}|\chi^{0}_{0}\rangle_n
,
\label{L1}
\end{eqnarray}
where the standard scalar product for the creation and
annihilation operators is assumed.
This Lagrangian is now written for fields with given spin which
are defined by $n$ chosen according to (\ref{eigenv}), (\ref{n})
\begin{eqnarray}
\pi|\chi^0_0\rangle_n=\bigl(n+(D-4)/2\bigr)|\chi^0_0\rangle_n,
&\qquad&
\pi|\chi^1_0\rangle_n=\bigl(n+(D-4)/2\bigr)|\chi^1_0\rangle_n
\end{eqnarray}
and the operator $K_n$ is the operator $K$ (\ref{K}) where
the following substitution is assumed be done
$h\to-\bigl(n+(D-4)/2\bigr)$
\begin{eqnarray}
K_n&=&
\sum_{k=0}^\infty\frac{1}{k!}
  \Bigl(\,
     |k\rangle{}\langle{}k|\,\,\,\,C(k,-n-\frac{D-4}{2})
\nonumber
\\&&
\qquad{}
-
     2d^+|k\rangle\langle{}k|\,d\,\,\,\,C(k+1,-n-\frac{D-4}{2})\,
  \,\Bigr).
\label{Kn}
\end{eqnarray}

Thus the operators $K_n$ depend on
the spin of the fields.
Note also that we can write $\Delta{}Q_n$ instead of
$\Delta{}Q_{\pi}$ in the equations of motion (\ref{EofM1}),
(\ref{EofM2}), in the Lagrangian (\ref{L1})
and in the gauge transformations (\ref{GT1})--(\ref{GTi2})
for fixed spin fields below.

The equations of motion (\ref{EofM1}), (\ref{EofM2}) and the
Lagrangian (\ref{L1}) are invariant under the gauge
transformations
\begin{eqnarray}
\delta|\chi^{0}_{0}\rangle_n
&=&
\Delta{}Q_{\pi}|\Lambda^{0}_{0}\rangle_n
 +
 \frac{1}{2}\bigl\{\tilde{T}_0,\eta_1^+\eta_1\bigr\}
 |\Lambda^{1}_{0}\rangle_n,
\label{GT1}
\\
\delta|\chi^{1}_{0}\rangle_n
&=&
\tilde{T}_0|\Lambda^{0}_{0}\rangle_n
 +\Delta{}Q_{\pi}|\Lambda^{1}_{0}\rangle_n
 .
\label{GT2}
\end{eqnarray}
which are reducible
\begin{align}
\delta|\Lambda^{(i)}{}^{0}_{0}\rangle_n
&=
\Delta{}Q_{\pi}|\Lambda^{(i+1)}{}^{0}_{0}\rangle_n
 +
 \frac{1}{2}\bigl\{\tilde{T}_0,\eta_1^+\eta_1\bigr\}
 |\Lambda^{(i+1)}{}^{1}_{0}\rangle_n,
&
|\Lambda^{(0)}{}^0_0\rangle_n=|\Lambda^0_0\rangle_n,
\label{GTi1}
\\
\delta|\Lambda^{(i)}{}^{1}_{0}\rangle_n
&=
\tilde{T}_0|\Lambda^{(i+1)}{}^{0}_{0}\rangle_n
 +\Delta{}Q_{\pi}|\Lambda^{(i+1)}{}^{1}_{0}\rangle_n,
&
|\Lambda^{(0)}{}^1_0\rangle_n=|\Lambda^1_0\rangle_n,
\label{GTi2}
\end{align}
with finite number of reducibility stages $i_{max}=n-1$ for spin
$s=n+1/2$. In Section~\ref{s:FF} we show that the Lagrangian
(\ref{L1}) is transformed to the Fang~\&~Fronsdal Lagrangian
\cite{Fang} in four dimensions after eliminating the auxiliary
fields.  So, we construct the Lagrangian for arbitrary fixed spin
fermionic fields using the BRST approach.
Now we turn to construction of Lagrangian describing propagation
of all half-integer spin fields simultaneously.

\section{Lagrangian for all half-integer spin fields}\label{LagrAll}

In this section we construct the Lagrangian which describes all
half-integer spin fields simultaneously, i.e. we construct the
Lagrangian in terms of the fields containing all half-integer
spins
\begin{eqnarray}
|\chi^i_0\rangle&=&\sum_{n=0}^{\infty}|\chi^i_0\rangle_n,
\qquad
i=0,1.
\label{fall}
\end{eqnarray}

As we mentioned above the operator $\pi$ commutes with the
operators $Q_{\pi}$, $\tilde{T}_0$,
$\bigl\{\tilde{T}_0,\eta_1^+\eta_1\bigr\}$ and moreover it
commutes with each term of these operators.
Therefore we can write all the operators $K_n$ in the
Lagrangians (\ref{L1}) in the same form for any spin.
This is done analogously to that when we transformed $Q_n$
into $Q_{\pi}$.
Namely, we stand all $h$ to the right (or to the left position)
in the expression for $K$ (\ref{K}) and substitute $\pi$ instead
of $-h$. As a result we have
\begin{eqnarray}
\label{Kpi}
K_{\pi}&=&
\sum_{n=0}^\infty\frac{1}{n!}
  \Bigl(\,
     |n\rangle{}\langle{}n|\,\,\,\,C(n,-\pi)
     -
     2d^+|n\rangle\langle{}n|\,d\,\,\,\,C(n+1,-\pi)\,
  \,\Bigr).
\end{eqnarray}
Thus we can substitute $K_{\pi}$ (\ref{Kpi}) instead of $K_n$
(\ref{Kn}) in the expression for the Lagrangian corresponding to
one fixed spin field (\ref{L1}).

Evidently that Lagrangian describing all half-integer spin
fields simultaneously should be a sum of all the Lagrangians
for each spin (\ref{L1})
\begin{eqnarray}
{\cal{}L}&=&\sum_{n=0}^{\infty}{\cal{}L}_n
=
\sum_{n=0}^{\infty}\Bigl[
{}_n\langle\chi^{0}_{0}|K_{\pi}\tilde{T}_0|\chi^{0}_{0}\rangle_n
+
\frac{1}{2}\,{}_n\langle\chi^{1}_{0}|K_{\pi}\bigl\{
   \tilde{T}_0,\eta_1^+\eta_1\bigr\}|\chi^{1}_{0}\rangle_n
\nonumber
\\&&\qquad{}
+
{}_n\langle\chi^{0}_{0}|K_{\pi}\Delta{}Q_{\pi}|\chi^{1}_{0}\rangle_n
+
{}_n\langle\chi^{1}_{0}|K_{\pi}\Delta{}Q_{\pi}|\chi^{0}_{0}\rangle_n
\Bigr]
.
\label{Lall-}
\end{eqnarray}

Now we rewrite this Lagrangian in terms of two concise fields
$|\chi_0^0\rangle$ and $|\chi_0^1\rangle$ (\ref{fall})
containing all half-integer spin fields.
Using that
\begin{math}
{}_n\langle\chi^i_0|\chi^{i'}_0\rangle_{n'}
\sim
\delta^{ii'}\delta_{nn'},
\end{math}
we transform each term in Lagrangian (\ref{Lall-}) as
\begin{eqnarray}
\sum_{n=0}^{\infty}
{}_n\langle\chi^{0}_{0}|K_{\pi}\tilde{T}_0|\chi^{0}_{0}\rangle_n
=
\Bigl(\sum_{n=0}^{\infty}{}_n\langle\chi_0^0|\Bigr)
K_{\pi}\tilde{T}_0
\Bigl(\sum_{n=0}^{\infty}|\chi_0^0\rangle_n\Bigr)
=
\langle\chi^{0}_{0}|K_{\pi}\tilde{T}_0|\chi^{0}_{0}\rangle
\end{eqnarray}
and find
\begin{eqnarray}
{\cal{}L}&=&
\langle\chi^{0}_{0}|K_{\pi}\tilde{T}_0|\chi^{0}_{0}\rangle
+
\frac{1}{2}\langle\chi^{1}_{0}|K_{\pi}\bigl\{
   \tilde{T}_0,\eta_1^+\eta_1\bigr\}|\chi^{1}_{0}\rangle
\nonumber
\\
&&\qquad{}
+
\langle\chi^{0}_{0}|K_{\pi}\Delta{}Q_{\pi}|\chi^{1}_{0}\rangle
+
\langle\chi^{1}_{0}|K_{\pi}\Delta{}Q_{\pi}|\chi^{0}_{0}\rangle
\label{Lall}
\end{eqnarray}
where we have used (\ref{fall}).
Thus we have constructed the Lagrangian describing propagation
of all half-integer spin fields simultaneously (\ref{Lall}), the
equations of motion which are derived from it are
(\ref{EofM1all}), (\ref{EofM2all}).

Let us turn to the gauge transformations for the fields
(\ref{fall}). Summing up the gauge transformation for the fields
of fixed spins (\ref{GT1}), (\ref{GT2}) over all $n$ we get the
gauge transformations for the fields containing all half-integer
spins (\ref{fall})
\begin{eqnarray}
\delta|\chi^{0}_{0}\rangle
&=&
\Delta{}Q_{\pi}|\Lambda^{0}_{0}\rangle
 +
 \frac{1}{2}\bigl\{\tilde{T}_0,\eta_1^+\eta_1\bigr\}
 |\Lambda^{1}_{0}\rangle,
\label{GT1all}
\\
\delta|\chi^{1}_{0}\rangle
&=&
\tilde{T}_0|\Lambda^{0}_{0}\rangle
 +\Delta{}Q_{\pi}|\Lambda^{1}_{0}\rangle
 .
\label{GT2all}
\end{eqnarray}
which are also reducible
\begin{align}
\delta|\Lambda^{(i)}{}^{0}_{0}\rangle
&=
\Delta{}Q_{\pi}|\Lambda^{(i+1)}{}^{0}_{0}\rangle
 +
 \frac{1}{2}\bigl\{\tilde{T}_0,\eta_1^+\eta_1\bigr\}
 |\Lambda^{(i+1)}{}^{1}_{0}\rangle,
&
|\Lambda^{(0)}{}^0_0\rangle=|\Lambda^0_0\rangle,
\label{GTi1all}
\\
\delta|\Lambda^{(i)}{}^{1}_{0}\rangle
&=
\tilde{T}_0|\Lambda^{(i+1)}{}^{0}_{0}\rangle
 +\Delta{}Q_{\pi}|\Lambda^{(i+1)}{}^{1}_{0}\rangle,
&
|\Lambda^{(0)}{}^1_0\rangle=|\Lambda^1_0\rangle,
\label{GTi2all}
\end{align}
where we introduced
\begin{eqnarray}
|\Lambda^{(j)}{}^i_0\rangle
&=&
\sum_{n=i+j+1}^{\infty}|\Lambda^{(j)}{}^i_0\rangle_n
\qquad
i=0,1,
\qquad
j=0,1,\ldots
.
\end{eqnarray}
Since the fields (\ref{fall}) contain infinite number of spins
and since the order of reducibility grows with the spin value,
then the order of reducibility of the gauge symmetry for fields
(\ref{fall}) will be infinite.

It should be noted that the procedure developed here for
constructing the Lagrangians both for fixed spin fields (\ref{L1})
and for all half-integer spin fields (\ref{Lall}) may be
used for constructing Lagrangians for bosonic fields as well.
Also this procedure may be generalized for constructing
Lagrangians with mixed symmetry tensor-spinor fields as it was
done in the bosonic case \cite{0101201}.

Now we turn to the reduction of (\ref{L1}) to the
Fang~\&~Fronsdal Lagrangian \cite{Fang}.

\section{Reduction to Fang~{\&}~Fronsdal Lagrangian}\label{s:FF}

In this Section we show that Lagrangian (\ref{L1}) is transformed to
the Fang~{\&}~Fronsdal Lagrangian \cite{Fang} in four dimensions
after elimination of the auxiliary fields.

Let us consider Lagrangian (\ref{L1}) with some fixed $n$.
In this case the gauge symmetry (\ref{GTi1}), (\ref{GTi2}) is
reducible with $i_{max}=n-1$.
We can write down the dependence of the fields and the gauge
parameters on the ghost fields explicitly. For the lowest gauge
parameters we have
\begin{eqnarray}
|\Lambda^{(n-1)}{}^{0}_{0}\rangle_n&=&(p_1^+)^{n-1}
\Bigl\{
{\cal{}P}_1^+|\omega\rangle_0
+
p_1^+|\omega_1\rangle_0
\Bigr\}
,
\\
|\Lambda^{(n-1)}{}^{1}_{0}\rangle_n&=&0.
\end{eqnarray}
Here we have taken into account that the gauge parameters are
the eigenvectors of the operator $\pi$ with the eigenvalue
$-\bigl(n+\frac{D-4}{2}\bigr)$ and that they have the ghost
numbers $-n$ and $-(n+1)$ respectively.
Let us recall that the subscripts at the state vectors are
associated with the eigenvalues of the corresponding state
vectors (\ref{chin}).

With the help of these parameters we can get rid of the
dependence on the ghost ${\cal{}P}_2^+$ in the parameters
$|\Lambda^{(n-2)}{}^{0}_{0}\rangle_n$.
(The parameter $|\Lambda^{(n-2)}{}^{1}_{0}\rangle_n$ has no
dependence on the ghost ${\cal{}P}_2^+$.)
We may go on and get rid of any dependence on the ghost
${\cal{}P}_2^+$ in all the fields and the parameters.
The restriction which appears on the fields and the parameters is
that they can depend on the ghost $p_1^+$ maximum in the first
power.
(They must be annihilated by the operator $q_1^2$.)
Due to this restriction the remain gauge symmetry is
reducible with the first stage reducibility.

Then we can get rid of the dependence on the ghost $\eta_2^+$ in
all the remain fields and the gauge parameters. The
restriction appeared is that the fields and the gauge parameters
must be annihilated by the operator
$L_2'\equiv{}L_2+(b^+b+h)b+d^+db+\eta_1{\cal{}P}_1+iq_1p_1$.

After this we get rid of the gauge transformation parameter
$|\Lambda^{1}_{0}\rangle_n$ with the help of the parameter
$|\Lambda^{(1)}{}^{0}_{0}\rangle_n$.
Now we write down the remain fields and the gauge
parameter explicitly
\begin{eqnarray}
|\chi^{0}_{0}\rangle_n&=&
|\Psi\rangle_n+\eta_1^+{\cal{}P}_1^+|\Psi_1\rangle_{n-2}
+q_1^+p_1^+|\Psi_2\rangle_{n-2}
+p_1^+\eta_1^+|\Psi_3\rangle_{n-2}
+q_1^+{\cal{}P}_1^+|\Psi_4\rangle_{n-2}
\nonumber
\\
&&\qquad{}
+q_1^+p_1^+\eta_1^+{\cal{}P}_1^+|\Psi_5\rangle_{n-4}
+q_1^{+2}p_1^+{\cal{}P}_1^+|\Psi_6\rangle_{n-4},
\\
|\chi^{1}_{0}\rangle_n&=&
{\cal{}P}_1^+|\chi\rangle_{n-1}
+p_1^+|\chi_1\rangle_{n-1}
+p_1^+\eta_1^+{\cal{}P}_1^+|\chi_2\rangle_{n-3}
+q_1^+p_1^+{\cal{}P}_1^+|\chi_3\rangle_{n-3},
\\
|\Lambda^{0}_{0}\rangle_n&=&
{\cal{}P}_1^+|\xi\rangle_{n-1}
+p_1^+|\xi_1\rangle_{n-1}
+q_1^+p_1^+{\cal{}P}_1^+|\xi_2\rangle_{n-3}
+p_1^+\eta_1^+{\cal{}P}_1^+|\xi_3\rangle_{n-3}
\end{eqnarray}
with the restriction
\begin{math}
L_2'|\chi^{0}_{0}\rangle_n
=L_2'|\chi^{1}_{0}\rangle_n
=L_2'|\Lambda^{0}_{0}\rangle_n
=0.
\end{math}
Here $|\Psi_i\rangle_k$, $|\chi_i\rangle_k$, $|\xi_i\rangle_k$
do not depend on the ghost fields.
Using the gauge transformations we first get rid of the fields
$|\Psi_2\rangle_{n-2}$, $|\Psi_4\rangle_{n-2}$,
$|\Psi_6\rangle_{n-4}$ after which
the gauge parameters are restricted by
\begin{math}
T_1'|\xi_1\rangle_{n-1}
=T_1'|\xi\rangle_{n-1}
=T_1'|\xi_2\rangle_{n-3}=0,
\end{math}
with $T_1'\equiv{}T_1-2(b^+b+h)d-d^+b$.
Now we can see that $|\chi_3\rangle_{n-3}=0$ and then
$|\Psi_5\rangle_{n-4}=0$ as the equation of motion.
Then we eliminate one after another the fields
$|\chi_2\rangle_{n-3}$,
$|\Psi_3\rangle_{n-2}$ and $|\chi_1\rangle_{n-1}$.
The new restrictions on the gauge parameters are
\begin{math}
T_0|\xi_3\rangle_{n-3}
=L_1|\xi_1\rangle_{n-1}
=T_0|\xi_1\rangle_{n-1}
=0.
\end{math}
The remain gauge freedom is enough to get rid of the dependence
on $b^+$ and $d^+$ in $|\Psi\rangle_n$, $|\Psi_1\rangle_{n-2}$ and
$|\chi\rangle_{n-1}$.
After this the remain equations of motion and the gauge
transformation are
\begin{align}
\label{e1}
&T_0|\Psi_0\rangle_n+L_1^+|\chi_0\rangle_{n-1}=0,
&
&T_1|\Psi_0\rangle_n=|\chi_0\rangle_{n-1},
\\
\label{e2}
&T_0|\chi_0\rangle_{n-1}-L_1|\Psi_0\rangle_n
  +L_1^+|\Psi_{10}\rangle_{n-2}=0,
&
&T_1|\chi_0\rangle_{n-1}=-2|\Psi_{10}\rangle_{n-2},
\\
\label{e3}
&T_0|\Psi_{10}\rangle_{n-2}+L_1|\chi_0\rangle_{n-1}=0,
&
&T_1|\Psi_{10}\rangle_{n-2}=0,
\\[2mm]
\label{gt0}
&\delta|\Psi_0\rangle_n=L_1^+|\xi_0\rangle_{n-1},
\\
&\delta|\chi_0\rangle_{n-1}=-T_0|\xi_0\rangle_{n-1},
&
&\delta|\Psi_{10}\rangle_{n-2}=L_1|\xi_0\rangle_{n-1}.
\end{align}
Here subscript $0$ means that the corresponding fields
and the gauge parameter do not depend on $b^+$ and $d^+$.
Besides, the gauge parameter is restricted as in the Fang and
Fronsdal theory $T_1|\xi_0\rangle_{n-1}=0$.
The equations which stand in the left column of
(\ref{e1})--(\ref{e3}) can be derived from the Lagrangian
\begin{eqnarray}
{\cal{}L}'&=&
{}_n\langle\Psi_0|\bigl\{
T_0|\Psi_0\rangle_n+L_1^+|\chi_0\rangle_{n-1}
\bigr\}
-{}_{n-1}\langle\chi_0|
  \bigl\{T_0|\chi_0\rangle_{n-1}-L_1|\Psi_0\rangle_n
     +L_1^+|\Psi_{10}\rangle_{n-2}\bigr\}
\nonumber
\\
&&\qquad{}
-{}_{n-2}\langle\Psi_{10}|\bigl\{T_0|\Psi_{10}\rangle_{n-2}
   +L_1|\chi_0\rangle_{n-1}\bigr\}.
\label{FFL0}
\end{eqnarray}
Using the restrictions on the fields which stand in the right column
of (\ref{e1}), (\ref{e2}) we can express the fields
$|\Psi_{10}\rangle_{n-2}$ and $|\chi_{0}\rangle_{n-1}$ through
$|\Psi_0\rangle_n$ and substitute them in the Lagrangian
(\ref{FFL0}).
As a result ones get the Lagrangian which is generalization of the Fang
and Fronsdal Lagrangian \cite{Fang} for arbitrary dimensional spacetime
\begin{eqnarray}
{\cal{}L}
&=&
{}_n\langle\Psi_0| \bigl\{
T_0-T_1^+T_0T_1-L_2^+T_0L_2
\nonumber
\\&&
\qquad\qquad\qquad{}
+T_1^+L_1+L_1^+T_1
+L_2^+L_1T_1+T_1^+L_1^+L_2
\bigr\}
|\Psi_0\rangle_n,
\label{FFL}
\end{eqnarray}
with the vanishing triple $\gamma$-trace
$(T_1)^3|\Psi_0\rangle_n=0$ and gauge transformation (\ref{gt0})
with the constrained gauge parameter $T_1|\xi_0\rangle_{n-1}=0$.

To see that the Lagrangian (\ref{FFL}) indeed coincides with the
Lagrangian of Fang and Fronsdal \cite{Fang} we calculate it
(\ref{FFL}) explicitly for an arbitrary spin field $s=n+1/2$
\begin{eqnarray}
|\Psi_0\rangle_n
&=&\frac{1}{\sqrt{n!}}\,\, a^{+\mu_1}\cdots
a^{+\mu_n}h_{\mu_1\cdots\mu_n}(x)|0\rangle.
\label{psi}
\end{eqnarray}
First we find that
\begin{align}
&
{}_n\langle\Psi_0|T_0|\Psi_0\rangle_n=
(-1)^n\,\bar{h}\,\slash\!\!\!p\,h,
\\
&
{}_n\langle\Psi_0|T_1^+T_0T_1|\Psi_0\rangle=_n
(-1)^n\,n\,\bar{h}'\,\slash\!\!\!p\,h',
&&
{}_n\langle\Psi_0|L_2^+T_0L_2|\Psi_0\rangle_n=
(-1)^n\,\frac{n(n-1)}{4}\,\bar{h}''\,\slash\!\!\!p\,h'',
\\
&
{}_n\langle\Psi_0|T_1^+L_1|\Psi_0\rangle_n=
-(-1)^n\,n\,\bar{h}\,p\cdot{}h,
&&
{}_n\langle\Psi_0|L_1^+T_1|\Psi_0\rangle_n=
{}_n\langle\Psi_0|T_1^+L_1|\Psi_0\rangle_n^*
\\
&
{}_n\langle\Psi_0|L_2^+L_1T_1|\Psi_0\rangle_n=
(-1)^n\frac{n(n-1)}{2}\bar{h}''p\cdot{}h',
&&
{}_n\langle\Psi_0|T_1^+L_1^+L_2|\Psi_0\rangle_n=
{}_n\langle\Psi_0|L_2^+L_1T_1|\Psi_0\rangle_n^*,
\end{align}
where we have used the notation of \cite{Fang}.
Substituting the found relations in (\ref{FFL}) ones get
\begin{eqnarray}
{\cal{}L}&=&(-1)^n\Bigl(
\bar{h}\slash\!\!\!ph+n\bar{h}'\slash\!\!\!ph'
-\frac{1}{4}n(n-1)\bar{h}''\slash\!\!\!ph''
\nonumber
\\
&&\qquad\qquad{}
-n\bigl(\bar{h}'p\cdot{}h+\mbox{H.c.}\bigr)
+\frac{1}{2}n(n-1)\bigl(\bar{h}''p\cdot{}h'+\mbox{H.c.}\bigr)
\Bigr).
\label{FFL2}
\end{eqnarray}
The Lagrangian (\ref{FFL2}) coincides in $D=4$ with the Lagrangian of
Fang and Fronsdal up to overall factor $(-1)^n$.
It can be treated as Fang-Fronsdal Lagrangian for arbitrary
$D$-dimensional space.

Thus we showed that Lagrangian (\ref{L1}) is reduced to the
Fang and Fronsdal Lagrangian (\ref{FFL}) with all necessary
conditions on the field and the gauge parameter.

We point out that from the Lagrangian (\ref{FFL}) we may get
one Lagrangian describing propagation of all half-integer spin
fields simultaneously.
Summing up the Lagrangians (\ref{FFL}) over all half-integer
spins and noticing that
\begin{math}
{}_n\langle\Psi_0|\Psi_0\rangle_{n'}\sim\delta_{nn'}
\end{math}
we get this Lagrangian in the form
\begin{eqnarray}
{\cal{}L}
&=&
\langle\Psi_0| \bigl\{
T_0-T_1^+T_0T_1-L_2^+T_0L_2
+T_1^+L_1+L_1^+T_1
+L_2^+L_1T_1+T_1^+L_1^+L_2
\bigr\}
|\Psi_0\rangle,
\label{FFLall}
\end{eqnarray}
where we used the notation
\begin{eqnarray}
|\Psi_0\rangle&=&\sum_{n=0}^{\infty}|\Psi_0\rangle_n.
\end{eqnarray}

In the next Section, we apply our procedure
for derivation a new Lagrangian for spin-5/2
field model, where, unlike Fang-Fronsdal Lagrangian, all
auxiliary fields stipulated by general Lagrangian construction, are taken
into account.

\section{Construction of the Lagrangian for field with
spin 5/2}\label{example}

In this Section
we show how the generic Lagrangian construction (\ref{L1}), given in
terms of abstract state vectors, is transformed to standard space-time
Lagrangian form. We explicitly
derive a Lagrangian for the field with spin $5/2$ which
contains the auxiliary fields and more gauge symmetries in compare with
Fang - Fronsdal Lagrangian. Of course, it can be reduced to
Fang-Fronsdal Lagrangian after partial gauge-fixing and putting $D=4$.
However, this new Lagrangian possesses the interesting properties, in
particular it has a reducible gauge symmetry.

Let us start.
Since $s=n+1/2=5/2$
we have $n=2$ and $h=-\frac{D}{2}$
(\ref{n}).
Then we first extract the ghost fields dependence of the fields
and
the gauge parameters
\begin{eqnarray}
|\chi^{0}_{0}\rangle_2&=&
|\Psi\rangle_2+\eta_1^+{\cal{}P}_1^+|\Psi_1\rangle_0
+q_1^+p_1^+|\Psi_2\rangle_0
+p_1^+\eta_1^+|\Psi_3\rangle_0
+q_1^+{\cal{}P}_1^+|\Psi_4\rangle_0,
\label{f00}
\\
|\chi^{1}_{0}\rangle_2&=&
{\cal{}P}_1^+|\chi\rangle_1
+p_1^+|\chi_1\rangle_1
+{\cal{}P}_2^+|\chi_4\rangle_0,
\label{f10}
\\
|\Lambda^{0}_{0}\rangle_2&=&
{\cal{}P}_1^+|\xi\rangle_1
+p_1^+|\xi_1\rangle_1
+{\cal{}P}_2^+|\xi_4\rangle_0,
\label{l00}
\\
|\Lambda^{1}_{0}\rangle_2&=&
p_1^+{\cal{}P}_1^+|\lambda\rangle_0
+(p_1^+)^2|\lambda_1\rangle_0,
\label{l10}
\\
|\Lambda^{(1)}{}^{0}_{0}\rangle_2&=&
p_1^+{\cal{}P}_1^+|\omega\rangle_0
+(p_1^+)^2|\omega_1\rangle_0.
\label{l(1)}
\end{eqnarray}
Here the ghost numbers of the fields and the gauge parameters
are also taken into account.
In the following we omit the subscripts at the state
vectors associated with the eigenvalues of the operator $\pi$
(\ref{chin}).

Substituting the fields in this concise form in the Lagrangian
(\ref{L1}) ones find
\begin{eqnarray}
{\cal{}L}_2&=&
\langle\Psi|K_2\bigl\{
T_0|\Psi\rangle+L_1^+|\chi\rangle
+iT_1^{+\prime}|\chi_1\rangle+L_2^{+\prime}|\chi_4\rangle
\bigr\}
\nonumber
\\&&{}
+\langle\Psi_1|K_2\bigl\{
-T_0|\Psi_1\rangle-2i|\Psi_3\rangle
-L_1|\chi\rangle+|\chi_4\rangle
\bigr\}
\nonumber
\\&&{}
+\langle\Psi_2|K_2\bigl\{
T_0|\Psi_2\rangle-2|\Psi_3\rangle
+T_1'|\chi_1\rangle-i|\chi_4\rangle
\bigr\}
\nonumber
\\&&{}
+\langle\Psi_3|K_2\bigl\{
2i|\Psi_1\rangle-2|\Psi_2\rangle+iT_0|\Psi_4\rangle
+iT_1'|\chi\rangle
\bigr\}
\nonumber
\\&&{}
+\langle\Psi_4|K_2\bigl\{
-iT_0|\Psi_3\rangle+iL_1|\chi_1\rangle
\bigr\}
\nonumber
\\&&{}
+\langle\chi|K_2\bigl\{
-T_0|\chi\rangle-i|\chi_1\rangle+L_1|\Psi\rangle
-L_1^+|\Psi_1\rangle-iT_1^{+\prime}|\Psi_3\rangle
\bigr\}
\nonumber
\\&&{}
+\langle\chi_1|K_2\bigl\{
i|\chi\rangle-iT_1'|\Psi\rangle+T_1^{+\prime}|\Psi_2\rangle
-iL_1^+|\Psi_4\rangle
\bigr\}
\nonumber
\\&&{}
+\langle\chi_4|K_2\bigl\{
L_2'|\Psi\rangle+|\Psi_1\rangle+i|\Psi_2\rangle
\bigr\},
\label{L1-1}
\end{eqnarray}
where we have used that the ghost fields commute with the
operator $K_n$ (\ref{Kn}).
Next we find the gauge transformations (\ref{GT1}), (\ref{GT2})
\begin{align}
\label{100}
&
\delta|\Psi\rangle=
L_1^+|\xi\rangle
+iT_1^{+\prime}|\xi_1\rangle
+L_2^{+\prime}|\xi_4\rangle,
&
\delta|\Psi_1\rangle&=
L_1|\xi\rangle-|\xi_4\rangle+i|\lambda\rangle,
\\
&
\delta|\Psi_2\rangle=
T_1'|\xi_1\rangle-i|\xi_4\rangle-|\lambda\rangle,
&
\delta|\Psi_3\rangle&=
L_1|\xi_1\rangle-T_0|\lambda\rangle-2i|\lambda_1\rangle,
\\
&
\delta|\Psi_4\rangle=-T_1'|\xi\rangle,
&
\delta|\chi\rangle&=
-T_0|\xi\rangle-2i|\xi_1\rangle-iT_1^{+\prime}|\lambda\rangle,
\\
&\delta|\chi_1\rangle=
T_0|\xi_1\rangle+L_1^+|\lambda\rangle
+2iT_1^{+\prime}|\lambda_1\rangle,
&
\delta|\chi_4\rangle&=-T_0|\xi_4\rangle+4|\lambda_1\rangle,
\label{103}
\end{align}
and the gauge for gauge transformations (\ref{GTi1}), (\ref{GTi2})
\begin{align}
\label{104}
&
\delta|\xi\rangle=-iT_1^{+\prime}|\omega\rangle,
&
\delta|\xi_1\rangle&=L_1^+|\omega\rangle
+2iT_1^{+\prime}|\omega_1\rangle,
&
\delta|\xi_4\rangle&=4|\omega_1\rangle,
\\
&
\delta|\lambda\rangle=-T_0|\omega\rangle-4i|\omega_1\rangle,
&
\delta|\lambda_1\rangle&=T_0|\omega_1\rangle.
\label{105}
\end{align}
in the concise form.

Now in order to derive Lagrangian (\ref{L1}) (or
(\ref{L1-1})) in component form we write the fields
$|\Psi_i\rangle$ and $|\chi_i\rangle$ entering into (\ref{f00})
and (\ref{f10}) explicitly
(taking into account the field's eigenvalues associated with the
operator $\pi$)
\begin{eqnarray}
|\Psi\rangle&=&\Bigl\{
\frac{1}{2}a^{+\mu}a^{+\nu}\psi_{\mu\nu}(x)
+d^+a^{+\mu}\psi_{\mu}(x)
+b^+\psi(x)
\Bigr\}|0\rangle,
\\
|\Psi_1\rangle&=&\psi_1(x)|0\rangle,
\qquad\qquad
\qquad\qquad
\qquad
|\Psi_2\rangle=\psi_2(x)|0\rangle,
\\
|\Psi_3\rangle&=&\psi_3(x)|0\rangle,
\qquad\qquad
\qquad\qquad
\qquad
|\Psi_4\rangle=\psi_4(x)|0\rangle,
\\
|\chi\rangle&=&\Bigl\{
a^{+\mu}\chi_\mu(x)+d^+\chi(x)
\Bigr\}|0\rangle,
\\
|\chi_1\rangle&=&\Bigl\{
a^{+\mu}\chi_{1\mu}(x)+d^+\chi_1(x)
\Bigr\}|0\rangle,
\qquad
|\chi_4\rangle=\chi_4(x)|0\rangle
\end{eqnarray}
and substitute them into (\ref{L1-1}).
As a result we get the Lagrangian (\ref{L1}) for the field with
spin $5/2$ in the explicit form
\begin{eqnarray}
{\cal{}L}&=&
-i\bar{\psi}_{\mu\nu}\Bigl\{
\frac{1}{2}\gamma^\sigma\partial_\sigma\psi_{\mu\nu}
+\partial_\mu\chi_\nu
-\gamma_\mu\chi_{1\nu}
+\frac{i}{2}\eta_{\mu\nu}\chi_4
\Bigr\}
\nonumber
\\
&&{}
-iD\bar{\psi}^\mu\Bigl\{
\gamma^\sigma\partial_\sigma\psi_\mu
-\partial_\mu\chi+\chi_{1\mu}-\gamma_\mu\chi_1
\Bigr\}
+
\frac{i}{2}D\bar{\psi}\Bigl\{
\gamma^\mu\partial_\mu\psi-2\chi_1+i\chi_4
\Bigr\}
\nonumber
\\
&&{}
+i\bar{\psi}_1\Bigl\{
\gamma^\mu\partial_\mu\psi_1-2\psi_3
-\partial^\mu\chi_\mu-i\chi_4
\Bigr\}
\nonumber
\\
&&{}
-
\bar{\psi}_2\Bigl\{
i\gamma^\mu\partial_\mu\psi_2+2\psi_3+\gamma^\mu\chi_{1\mu}
-D\chi_1+i\chi_4
\Bigr\}
\nonumber
\\
&&{}
+\bar{\psi}_3\Bigl\{
2i\psi_1-2\psi_2+\gamma^\mu\partial_\mu\psi_4
-i\gamma^\mu\chi_\mu+iD\chi
\Bigr\}
-
\bar{\psi}_4\Bigl\{
\gamma^\mu\partial_\mu\psi_3+\partial^\mu\chi_{1\mu}
\Bigr\}
\nonumber
\\
&&{}
-i\bar{\chi}^\mu\Bigl\{
\gamma^\sigma\partial_\sigma\chi_\mu-\chi_{1\mu}
+\partial^\nu\psi_{\mu\nu}+\partial_\mu\psi_1
-\gamma_\mu\psi_3
\Bigr\}
-
iD\bar{\chi}\Bigl\{
\gamma^\mu\partial_\mu\chi+\chi_1
-\partial^\mu\psi_\mu+\psi_3
\Bigr\}
\nonumber
\\
&&{}
-
i\bar{\chi}^\mu_1\Bigl\{\chi_\mu+\gamma^\nu\psi_{\mu\nu}
-D\psi_\mu-i\gamma_\mu\psi_2+i\partial_\mu\psi_4
\Bigr\}
+
iD\bar{\chi}_1\Bigl\{\chi-\gamma^\mu\psi_\mu
+\psi-i\psi_2\Bigr\}
\nonumber
\\
&&{}
+
i\bar{\chi}_4\Bigl\{
\frac{1}{2}\psi^\mu{}_\mu-\frac{D}{2}\psi+\psi_1+i\psi_2
\Bigr\}.
\label{L5-2}
\end{eqnarray}
Here $D$ is dimension of the space-time, $\psi_{\mu\nu}$ is
basic spin 5/2 field and all other fields are auxiliary.
In order to write the gauge transformations in the explicit form
we write the gauge fields $|\xi_i\rangle$ and
$|\lambda_i\rangle$ entering into (\ref{l00}) and (\ref{l10}) as
follows
(also taking into account the gauge parameters's eigenvalues
associated with the operator $\pi$)
\begin{align}
&
|\xi\rangle=\bigl\{a^{+\mu}\xi_\mu(x)+d^+\xi(x)\bigr\}|0\rangle,
&
|\xi_1\rangle&=\bigl\{a^{+\mu}\xi_{1\mu}(x)+d^+\xi_1(x)\bigr\}|0\rangle,
\\
&|\xi_4\rangle=\xi_4(x)|0\rangle,
\\
&|\lambda\rangle=\lambda(x)|0\rangle,
&|\lambda_1\rangle&=\lambda_1(x)|0\rangle,
\end{align}
and substitute them into (\ref{100})--(\ref{103}). As a result we
get the gauge transformations for the field associated with spin
$5/2$ in the explicit form
\begin{eqnarray} \label{115}
\delta\psi_{\mu\nu}&=& -i(\partial_\mu\xi_\nu+\partial_\nu\xi_\mu)
+i(\gamma_\mu\xi_{1\nu}+\gamma_\nu\xi_{1\mu})
+\eta_{\mu\nu}\xi_4,
\\
\delta\psi_\mu&=&
-i\partial_\mu\xi+i\xi_{1\mu}-i\gamma_\mu\xi_1,
\qquad
\qquad
\quad
\delta\psi=2i\xi_1+\xi_4,
\\
\delta\psi_1&=&
i\partial^\mu\xi_\mu-\xi_4+i\lambda,
\qquad
\qquad\qquad
\quad
\delta\psi_2=
-\gamma^\mu\xi_{1\mu}+D\xi_1-i\xi_4-\lambda,
\\
\delta\psi_3&=&
i\partial^\mu\xi_{1\mu}+i\gamma^\mu\partial_\mu\lambda-2i\lambda_1,
\qquad
\qquad
\delta\psi_4=\gamma^\mu\xi_\mu-D\xi,
\\
\delta\chi_\mu&=&
i\gamma^\sigma\partial_\sigma\xi_\mu-2i\xi_{1\mu}-i\gamma_\mu\lambda,
\qquad
\qquad
\delta\chi=
-i\gamma^\mu\partial_\mu\xi-2i\xi_1-i\lambda,
\\
\delta\chi_{1\mu}&=&
-i\gamma^\sigma\partial_\sigma\xi_{1\mu}-i\partial_\mu\lambda
+2i\gamma_\mu\lambda_1,
\qquad
\delta\chi_1=
i\gamma^\mu\partial_\mu\xi_1+2i\lambda_1,
\\
\delta\chi_4&=&i\gamma^\mu\partial_\mu\xi_4+4\lambda_1.
\label{121}
\end{eqnarray}

Finally we get in the explicit form the gauge for gauge
transformations (\ref{104}), (\ref{105}).
Writing the gauge for gauge parameters as
\begin{eqnarray}
|\omega\rangle&=&\omega(x)|0\rangle,
\qquad
\qquad
|\omega_1\rangle=\omega_1(x)|0\rangle.
\end{eqnarray}
ones find (\ref{104}), (\ref{105}) in the explicit form
\begin{align}
\label{123}
&\delta\xi_\mu=-i\gamma_\mu\omega,&\delta\xi&=-i\omega,
\\
&\delta\xi_{1\mu}=-i\partial_\mu\omega+2i\gamma_\mu\omega_1,
&\delta\xi_1&=2i\omega_1,
&\delta\xi_4&=4\omega_1,
\\
&\delta\lambda=i\gamma^\mu\partial_\mu\omega-4i\omega_1,
&\delta\lambda_1&=-i\gamma^\mu\partial_\mu\omega_1.
\label{125}
\end{align}
Thus, following the general procedure described in Section~\ref{Lagr} we
have constructed the Lagrangian (\ref{L1}), the gauge
transformations (\ref{GT1}), (\ref{GT2}) and the gauge for gauge
transformations (\ref{GTi1}), (\ref{GTi2}) for the field model of
spin $5/2$ in the explicit form (\ref{L5-2}),
(\ref{115})--(\ref{121}), (\ref{123})--(\ref{125})
respectively. Unlike Fang-Fronsdal construction, we obtained the
Lagrangian containing all proper set of auxiliary fields.

\section{Summary}\label{Summary}

We have developed
the new BRST approach to derivation of Lagrangians for fermionic
massless higher spin models in arbitrary dimensional Minkowski
space.
We investigated
the superalgebra generated by the constraints which are
necessary to define an irreducible massless half-integer spin
representation of Poincare group and constructed the
corresponding BRST charge.
We found
that the model is reducible gauge theory and the order of
reducibility linearly grows with the value of spin.
It is shown
that this BRST charge generates the correct Lagrangian dynamics
for fermionic fields of any value of spin.
We construct
Lagrangians in the concise form for the fields of any fixed spin
in arbitrary space-time dimension and show that our Lagrangians
are reduced to the Fang-Fronsdall Lagrangians after partial
gauge-fixing.
As an example
of general scheme we obtained the Lagrangian and the gauge
transformations for the field of spin $5/2$ in the explicit
form without any gauge fixing.

The main results of the paper are given by the relations
(\ref{L1}),
where Lagrangian for the field with arbitrary half-integer spin is
constructed,
and
(\ref{GT1})--(\ref{GTi2})
where the gauge transformations for the fields and the gauge
parameters are written down.
In the case when ones consider all half-integer spin fields
together, the analogous relations are (\ref{Lall}) for the
Lagrangian and (\ref{GT1all})--(\ref{GTi2all}) for the gauge
transformations.
Our formulation does not impose any off-shell constraints on the
fields and the gauge parameters\footnote{The possibility to
formulate a higher spin field theory without restrictions on
traces of the fields and the gauge parameters was considered in
\cite{Francia}}
(see the discussion of this
point in \cite{0311257}).

The procedure for Lagrangian construction developed here for
higher spin massless fermionic field can be also applied to bosonic
higher spin massless theories and leads to the same results as in
\cite{9803207}. There are several possibilities for extending our
approach.  This approach may be applied to Lagrangian construction for
mixed symmetry tensor-spinor fields (see \cite{0101201} for
corresponding bosonic case), for Lagrangian construction for fermionic
fields in AdS background, for massive higher spin fields using the
dimensional reduction and for supersymmetric higher spin models.

\section*{Acknowledgements}
I.L.B. is very grateful to X. Berkaert, M. Grigoriev, M.
Tsulaia, M.A. Vasiliev for fruitful discussions.
We are thankful to A. Sagnotti and W. Siegel for useful
comments.
This work was supported in part by
the INTAS grants, projects
INTAS-03-51-6346
and
INTAS-00-00254,
The work of I.L.B. and V.A.K. was also supported by
the RFBR grant, project No.\ 03-02-16193,
the joint RFBR-DFG grant, project No.\ 02-02-04002,
the DFG grant, project No.\ 436 RUS 113/669,
the grant for LRSS, project No.\ 1252.2003.2
and
the grant PD02-1{.}2-94 of Russian Ministry of Education and
Science.
I.L.B. and V.A.K. are thankful the Humboldt-Universit\"at zu
Berlin, where part of this work was done
and  D. L\"ust for warm hospitality.

\end{document}